# Si-Ga$_2$O$_3$/p-GaN epitaxial heterostructure based self-powered and visible-blind UV photodetectors with fast and electrically tuneable response time

Ajoy Biswas, Amandeep Kaur, Bhabani Prasad Sahu, Sushantika Saha, Umakanta Patra, Rupa Jeena, Pradeep Sarin, and Subhabrata Dhar*

Department of Physics, Indian Institute of Technology Bombay, Powai, Mumbai 400076, India

Corresponding author email: dhar@phy.iitb.ac.in

**Abstract:**

n-Ga$_2$O$_3$/p-GaN heterojunction based photodetector devices are fabricated on Si-doped ($\bar{2}$01) $\beta$-Ga$_2$O$_3$ epitaxial layers grown by pulsed laser deposition (PLD) technique on p-type c-GaN/sapphire templates. These devices demonstrate the ability to act as highly efficient self-powered visible blind UV-photodetectors with fast response time. It has been found that the optimum performance of the detector in terms of its responsivity, detectivity and response time could be achieved by adjusting the Si doping level and the thickness of the Ga$_2$O$_3$ layer. Our best performing device showing the peak responsivity and detectivity of 56.8 mA/W and 3 × 10$^{12}$ Jones, respectively, is achieved for 660 nm thick Ga$_2$O$_3$ layer with Si-concentration of 8 × 10$^{18}$ cm$^{-3}$. Moreover, as low as a few nW of optical signal can be sensed by the detector. The response time of the detector is found to be only a few tens of nanoseconds, which highlights their potential for application in ultrafast detection of UV light. These devices also exhibit a slower component of photoresponse with a timescale of a few tens of milliseconds. Interestingly, the time-scale of the slower response can be prolonged by several orders of magnitude through enhancing the applied reverse bias. Such an electrical tuneability of the response time is highly desirable for neuromorphic device applications.

## I.    Introduction

Regular monitoring of the harmful UV radiation level in atmosphere is  becoming increasingly important due to its gradual rise as a result of depletion of ozone layer in the upper atmosphere caused by industrial activities.[1,2] Visible blind UV photodetectors should be best suited for this purpose. The demand for UV detectors with high detectivity and fast response time also comes from defence industries as such detectors can be used in missile detection-warning systems and fighter jet tracking.[2–4] UV detectors are also needed for space science, water purification and biomedical applications.[3–6] Often the photodetectors are operated under external bias. A more cost-effective approach is to use depletion region (p-n or Schottky junction) as the active region that can efficiently separate the photogenerated electrons and holes due to the photo-voltaic (PV) effect, which does not require external biasing.[7,8] The PV based detectors are also reported to be faster than those operating under applied bias.[9] This self-powered mode of operation enhances the potential for large-scale application of these detectors. n-ZnO/p-GaN heterojunction based self-powered UV detectors with fast response time are  reported by several group.[9–13]  However, these detectors can efficiently sense only the UVA band (315 – 400 nm).[9–12] Wider band-gap semiconductors are needed for detecting deeper UV radiations. $\beta$-Ga$_2$O$_3$ with a bandgap of ~4.9eV has emerged as a strong candidate for deep UV detection.[14–16] Stable n-type doping is achievable in this material. In fact, as grown $\beta$-Ga$_2$O$_3$ layer exhibits n-type behaviour, which  arises from the growth induced unintentional inclusion of oxygen vacancies acting as shallow donors. Controllable n-type doping has also been achieved in the material with different group IV elements such as Si, Sn and Ge.[17–21] However, achieving p-type doping in this material is hard.[22] Growing heterojunctions of n-type Ga$_2$O$_3$ with p-type GaN, NiO and Si, can be an alternative way to circumvent the challenge of p-type doping in Ga$_2$O$_3$.[23–26] Note that GaN is a wide bandgap semiconductor with a direct gap of ~3.4 eV.[27,28] The material is extensively used for UV and visible optoelectronics as well as high-power, high-frequency and high-temperature electronics. Moreover, (0001) plane of GaN offers as low as 4.4% lattice mismatch with ($\bar{2}$01) plane of $\beta$-Ga$_2$O$_3$[29,30], making the former a good substrate  for the epitaxial growth of the latter.[23,31–33] Furthermore, both GaN and $\beta$-Ga$_2$O$_3$ has good thermal and chemical stability.[34–36] A n-type $\beta$-Ga$_2$O$_3$/p-GaN heterojunction based photodetectors are thus expected to cover a broad wavelength region ranging from 250 to 400 nm. Blindness to the visible light should also help in enhancing the UV detectivity of these devices.

In recent years, though there are significant efforts concentrated on developing self-powered UV photodetectors, the reports on n-type $\beta$-Ga$_2$O$_3$/p-GaN heterojunction based UV photodetectors are not too many. Li et al. have studied the performance of n-Ga$_2$O$_3$/p-GaN based self-powered UV photodetectors.[23] Unintentionally n-type (background) doped $\beta$-Ga$_2$O$_3$ epitaxial films are grown on p-type (0001)GaN templates using PLD growth technique and the detectors based on these heterojunctions are found to show  maximum responsivity  of ~28.5 mA/W and rise/decay time of 0.14/0.07 s under 254 nm  light exposure.[23] Guo et al. have also studied n-type $\beta$-Ga$_2$O$_3$/p-GaN heterojunction based UV photodetectors, where Ga$_2$O$_3$ layer  is intentionally n-type doped with Sn. These devices show the maximum responsivity of ~3 A/W with fast and slow response times of

18 ms and 148 ms, respectively, under 254 nm illumination.[19] Interestingly, same order of magnitude high responsivity has also been reported in amorphous-$Ga_2O_3$/p-GaN based photodetectors.[37] More recently, Chen *et al.* reported self-powered Si: $\beta$-$Ga_2O_3$/i-$\beta$-$Ga_2O_3$/p-GaN based n-type/intrinsic/p-type (PIN) photodetectors. These PIN photodetectors are shown to exhibit higher responsivity of 72 mA/W with rise/decay time of 7/19 ms as compared to only hetero-p-n-junction Si: $\beta$-$Ga_2O_3$/p-GaN photodetectors, which show the maximum responsivity of 14 mA/W with rise/decay time of 37/42 ms under 260 nm light illumination.[17] Very high responsivity has also been reported in Ta: $\beta$-$Ga_2O_3$/i- $\beta$-$Ga_2O_3$/p-GaN based PIN UV photodetectors.[38]

Note that in case of UV photodetectors based on n-type $Ga_2O_3$ layers grown upon p-type GaN templates, the incorporation of n-type dopants in $Ga_2O_3$ lattice can affect the crystalline quality of the film and hence its optical, transport and photo-transport properties. Further, the depletion width in the n-$Ga_2O_3$ side, which depends on the doping concentration, plays a crucial role in governing the performance of such devices. Another important factor is the light absorption length-scale (absorption coefficient) of $Ga_2O_3$. The performance of such a device is expected to be maximized when the thickness of the n-$Ga_2O_3$ layer matches with the absorption length-scale as well as the depletion width. It is thus important to study the device performance of n-$Ga_2O_3$/p-GaN heterojunctions as a function of the doping concentration and the thickness of the n-$Ga_2O_3$ layer to obtain the optimum device parameters.

Here, Si-doped ($\bar{2}01$) $\beta$-$Ga_2O_3$ films are grown using PLD technique on p-type c-GaN/sapphire substrates. These heterojunctions are found to show rectifying current-voltage characteristics demonstrating p-n junction behavior. Effect of Si doping and the thickness of the $Ga_2O_3$ layer on the photodetection properties such as the responsivity, detectivity and the time response of these devices is systematically investigated. It has been observed that the heterojunction can operate as highly efficient visible blind UV-detectors under zero-bias condition. The performance can be maximized at certain optimum $Ga_2O_3$ layer thickness and Si-concentration. Here, the best performing device, which shows the peak responsivity and detectivity of 56.8 mA/W and $3 \times 10^{12}$ Jones, respectively, could be achieved at a Si-concentration of $8 \times 10^{18}$ cm$^{-3}$ and $Ga_2O_3$ thickness of 660 nm. Further, the detector is found to sense a signal as feeble as a few nW. Moreover, it is found to respond with a time-scale of a few tens of nanoseconds, which underscores the prospect of these heterojunctions for applications in ultrafast detection of UV light. These devices also exhibit a slower component of photoresponse, which is of the order of a few tens of milliseconds. Moreover, this component is shown to be increased by several orders of magnitude through increase of the applied reverse bias. As far as the neuromorphic device applications are concerned, the ability to control the response time through electrical means is highly interesting.

## II. Experimental Section

All $Ga_2O_3$ films were grown on Mg-doped p-type (0001)GaN/c-sapphire templates using pulse laser deposition (PLD) technique. GaN/sapphire templates (commercially procured from M/s TDI Corporation, USA) were consisted of ~ 4.5 μm thick [0001] oriented GaN epitaxial layers grown on c-sapphire (thickness: 430 μm) substrates using the hydride vapor phase epitaxy (HVPE) technique. In the PLD setup, a KrF excimer laser of 248 nm wavelength with a pulse duration of 30 ns was used to ablate the $Ga_2O_3$ target. The target was placed at a distance of 4.5 cm from the substrate. Pulse energy density and frequency of the laser were set at the optimized values of 2 J/cm$^2$ and 5 Hz, respectively, for all depositions. The base pressure of the chamber was below $1 \times 10^{-5}$ mbar. Before loading in the chamber, the substrate was cleaned by ultrasonication in trichloroethylene, acetone, 2-propanol for 5 min each and then immersed in dilute hydrofluoric acid (HF:H$_2$O ratio 1:10) for 1 min before rinsing in deionized water and drying under N$_2$ flow. Apart from the $Ga_2O_3$ pellet, three other Si-doped $Ga_2O_3$ pellets with Si-content of 0.1 at.%, 0.4 at.% and 1 at.% were used as targets. In all cases, oxygen (purity 99.999%) pressure was maintained at $1 \times 10^{-2}$ mbar and the growth temperature ($T_G$) was fixed at 750°C. After the growth, samples were cooled naturally to 100°C keeping the oxygen pressure intact. A series of samples with $Ga_2O_3$ film-thickness ranging between 50 to 1000 nm were grown for each of the Si-doped and -undoped targets by controlling the number of laser pulses falling on the target in the range of 500 to 10000. Series of samples deposited using 0 at.% (Si-undoped), 0.1 at.%, 0.4 at.% and 1 at.% Si-contained $Ga_2O_3$ pellets. The thickness of the samples were measured by ellipsometry and cross-sectional scanning electron microscopy (SEM) techniques.

Surface morphology of the grown layers was studied by atomic force microscopy (AFM) and field emission scanning electron microscopy (FEG-SEM). The crystalline properties of the films were investigated by x-ray diffraction (XRD) technique. $\omega - 2\theta$ and $\phi$ scans were recorded using a Rigaku Smart Lab diffractometer equipped with a 9 kW rotating Cu anode as a source for the Cu $K_\alpha$ ($\lambda$ = 1.5406 Å) x-ray radiation. Secondary ion mass spectroscopy (SIMS) analysis of the $Ga_2O_3$ films was performed to estimate the actual Si incorporation in the films. A commercially procured Si-doped $Ga_2O_3$ sample with known Si-content was used as a reference standard. In SIMS, a beam of O$_2^+$ ions with energy 5 keV and beam-current 400 nA was used to sputter the films.

A series of photodetector devices are fabricated on the $Ga_2O_3$ layers of different thicknesses grown with each of the Si-containing (including the Si-undoped) targets (see Table S1 in the supplementary material). Here, we highlight four devices (D7, D16, D26 and D33), one each from the four batches of Si-doped $Ga_2O_3$ layers showing the highest overall responsivity. The values of thickness and the actual Si-concentration (as measured by SIMS) of the $Ga_2O_3$ layer for these devices are listed in Table 1. Multiple metal contacts(diameter 1 mm) of Ti/Au(20/100 nm) and Ni/Au(30/60 nm) were deposited on $Ga_2O_3$ and GaN sides in each of these devices, respectively, using DC sputtering technique. Note that as deposited Ni/Au contacts on GaN substrate showed Ohmic nature (see Fig. S2(a) in the supplementary material). Because of the presence of underlaying p-GaN template, the behavior of the Ti/Au contacts on the Ga$_2$O$_3$ film was difficult to be examined. Ti/Au contacts were thus deposited on a Ga$_2$O$_3$ film grown directly on a c-sapphire substrate. These contacts were found to exhibit Ohmic behaviour (see Fig. S2(b) in the supplementary material). Current-voltage (I-V) measurements were carried out using a Keithley made 6487 picoammeter voltage source. It should be mentioned that all devices including the undoped ones exhibit rectifying current-voltage (I-V) characteristics, which implies the formation of p-n junction in all cases (see Fig. S2(c) in the supplementary material). However, the current at any given forward bias was found to be higher in devices with Si-doped Ga$_2$O$_3$ layers as compared to those with undoped Ga$_2$O$_3$. These findings suggested that in all cases, Ga$_2$O$_3$ layers were n-type doped. While in Si-doped Ga$_2$O$_3$ layers, the source could be understood as the Si-incorporation itself. Undoped Ga$_2$O$_3$ layers are unintentionally doped likely to be due to the formation of oxygen vacancies during growth.

For photoconductivity measurements, a 150 W Xenon white-light source and monochromator assembly was used to select the wavelength of the light that was focused on an area near to the metal contact in the Ga$_2$O$_3$ side of the device. To measure the temporal response, the device was illuminated with 254 nm light of power 67 µWcm$^{-2}$ after passing through a mechanical chopper. A 266 nm pulsed laser with a pulse width of 250 ps and a repetition rate of 8 kHz was also used to study the ultrafast response characteristics of these devices. The photocurrent signal was recorded using an oscilloscope with a bandwidth of 2.5 GHz. The intensity of the monochromatic light is measured by a calibrated power meter from Thorlabs.

### III. Results and Discussion

Figure 1(a) shows the out-of-plane $\omega - 2\theta$ XRD scans of different Si-doped Ga$_2$O$_3$ films grown for the pulse counts of 2000 on *c*-GaN/Sapphire templates. The profile for the *c*-GaN/Sapphire template is also presented in the figure. The peaks appearing at $2\theta$ values of 34.67° and 72.98° corresponds, respectively, to (0002) and (0004) planes of GaN. While 41.83° and 64.86° peaks can be assigned to (0006) and (0009) planes of sapphire. Apart from these features, a set of additional reflections appearing at 18.90°, 38.11°, 58.73° and 81.70° are evident in these samples, which can be attributed to the ($\bar{2}$01), ($\bar{4}$02), ($\bar{6}$03) and ($\bar{8}$04) planes of monoclinic ($\beta$-) phase of Ga$_2$O$_3$. Presence of mainly ($\bar{2}$01) and its higher order reflections of $\beta$-Ga$_2$O$_3$ implies that the layers are grown along [$\bar{2}$01]-direction in these samples. To study the in-plane epitaxial relationship between $\beta$-Ga$_2$O$_3$ and GaN, $\phi$ scans for ($\bar{4}$01) plane of $\beta$-Ga$_2$O$_3$ at a $\chi$ value 23.44° and (10$\bar{1}$2) plane of GaN at a $\chi$ value 43.1653° are recorded over a wide angular range as shown in Figure 1b for the undoped $\beta$-Ga$_2$O$_3$ sample. Evidently, in both the cases, six equally spaced peaks are observed. This is expected for GaN as (0001) plane of it has a six-fold symmetry. However, $\beta$-Ga$_2$O$_3$ has a monoclinic structure, it does not have six-fold symmetry. As the oxygen atoms on the ($\bar{2}$01) plane of $\beta$-Ga$_2$O$_3$ bonds with hexagonally arrange nitrogen atoms on the (0001) plane of GaN to form a distorted hexagon with [010] direction of $\beta$-Ga$_2$O$_3$ lies almost parallel to <11$\bar{2}$0> plane of GaN.[29] This is the reason why the $\phi$ scan of ($\bar{4}$01) $\beta$-Ga$_2$O$_3$ plane shows six equidistant peaks.

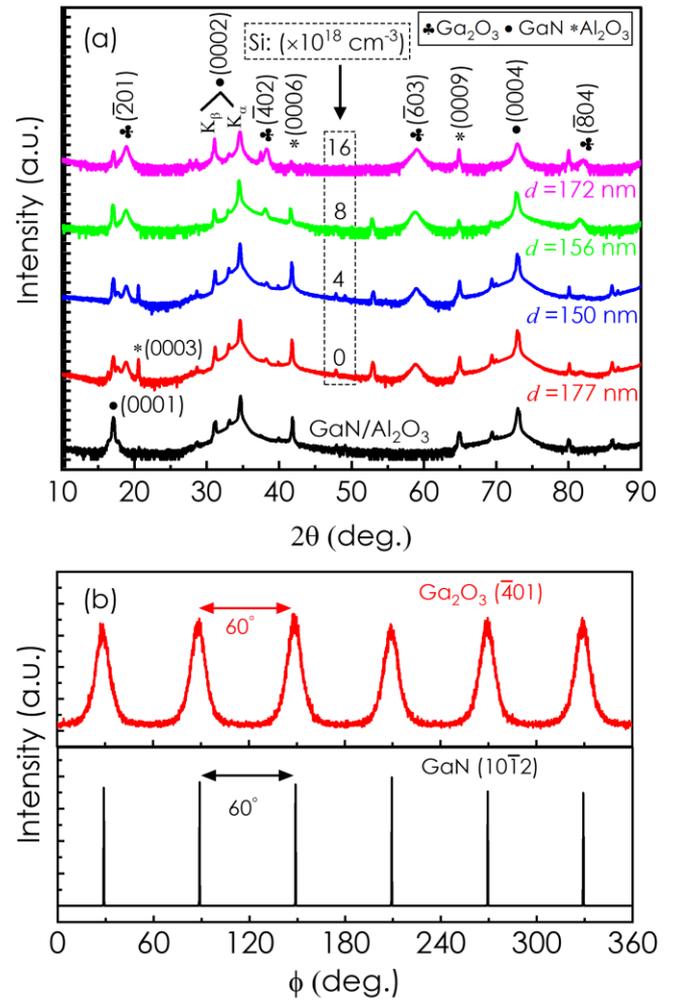

**FIG. 1.** (a) $\omega - 2\theta$ scans for samples with different Si-contents in Ga$_2$O$_3$. (b) $\phi$ scans for ($\bar{4}$01) plane of $\beta$-Ga$_2$O$_3$ and (10$\bar{1}$2) plane of GaN of the undoped sample.

Figure 2 shows the AFM images of undoped Ga$_2$O$_3$ film and three different Si-doped Ga$_2$O$_3$ films grown for the pulse counts of 2000. A continuous and smooth deposition is evident in all cases. Moreover, the surface morphology appears grainy for these layers. The grain-size seems to increase with the Si-content. AFM scans are recorded at various parts of the samples to obtain the average root-mean-square (RMS) roughness of these films. RMS roughness is found to be 0.69, 0.63, 0.88 and 1.04 nm, respectively, for the samples with Si concentrations of 0, $4 \times 10^{18}$, $8 \times 10^{18}$ and $1.6 \times 10^{19}$ cm$^{-3}$ [see the inset of Figure 2(d)]. This suggests an increase of the average grain size with the Si-content. SEM surface and cross-sectional images of these layers also reveal continuous and smooth depositions of the Ga$_2$O$_3$ layers with thickness ranging between 150 to 170 nm (see Fig. S3 and Table S1 in the supplementary material). Increase of the average size of the grains with the Si-content is also noticeable from the SEM images as well.

**Table 1.** Structural details and various performance parameters for our n-Ga$_2$O$_3$/p-GaN based self-powered photodetectors under 254 nm light illumination

| PD | Thickness of Ga$_2$O$_3$ layer $d$ [nm] | Si-content in targets [at.%] | Si-content in films [cm$^{-3}$] | Responsivity $R$ [mA/W] | Rise/decay time [ms] | Detectivity $D$ [Jones] | LDP [nW] |
|---|---|---|---|---|---|---|---|
| D7 | 670 | 0 | 0 | 2 | 6/4 | $1.1 \times 10^{11}$ | 10.9 |
| D16 | 660 | 0.1 | $4 \times 10^{18}$ | 14.9 | 4/6 | $8.1 \times 10^{11}$ | 3.02 |
| D26 | 660 | 0.4 | $8 \times 10^{18}$ | 36.5 | 6/10 | $1.9 \times 10^{12}$ | 1.31 |
| D33 | 490 | 1.0 | $1.6 \times 10^{19}$ | 10.1 | 18/13 | $5.2 \times 10^{11}$ | 12.2 |

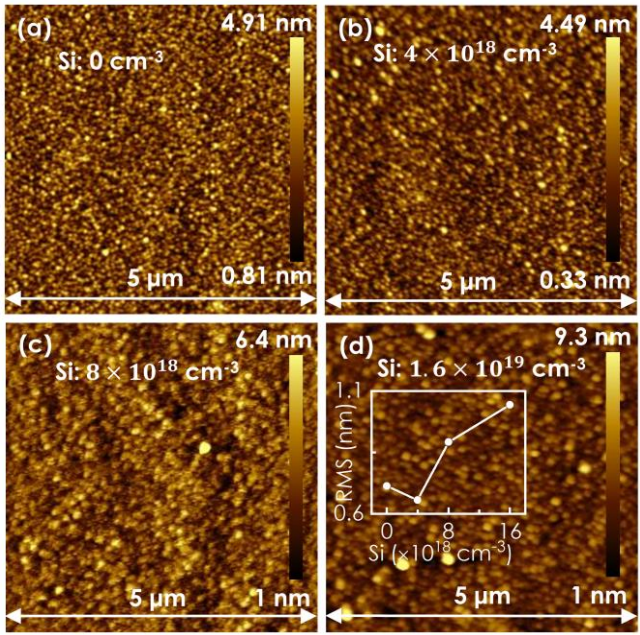

**FIG. 2.** AFM images for the samples with Si concentrations of (a) 0, (b) $4 \times 10^{18}$, (c) $8 \times 10^{18}$ and (d) $1.6 \times 10^{19}$ cm$^{-3}$. Inset of (d): RMS roughness as a function of Si concentration.

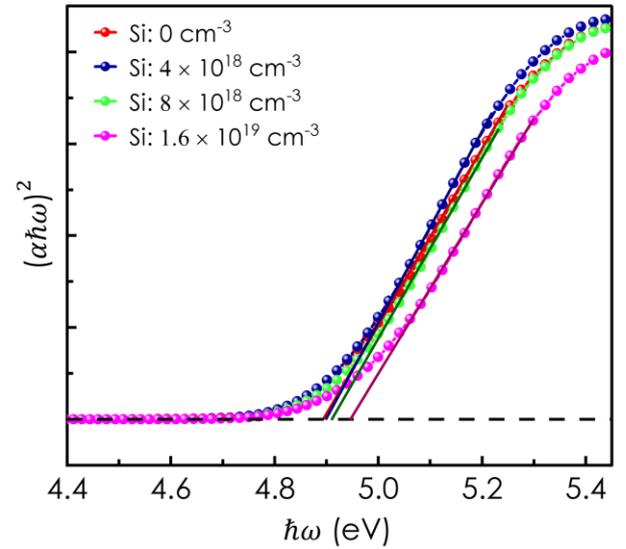

**FIG. 3.** Tauc plots [$(\alpha\hbar\omega)^2$ versus $\hbar\omega$] for different Si-doped Ga$_2$O$_3$ layers grown on c-sapphire substrates.

Figure 3 compares the photon energy dependence of the absorption coefficient $\alpha(\hbar\omega)$ [Tauc plots] measured using transmission spectroscopy for different Si-doped samples grown on double side polished c-sapphire substrates at the same growth conditions. The band gap estimated from the linear extrapolation of the data (shown in the figure) is found to range between 4.89 to 4.94 eV for these samples. A clear blue shift of the band-gap in the sample with Si content of $1.6 \times 10^{19}$ cm$^{-3}$ as compared to other samples might indicate a substantial enhancement of free electron population in this film due to high level of doping. High carrier concentration can push the fermi level into the conduction band leading to an increase of the optical-gap (Burstein–Moss effect).[39]

Figure 4(a) depicts the schematically the device structure. Figure 4(b)-(e) compare the current-voltage characteristics of the devices grown using 0.0 at.% (D7), 0.1 at.% (D16), 0.4 at.% (D26) and 1.0 at.% (D33) Si-containing targets (see Table 1), respectively, under dark and 254 nm light illuminated conditions. In all cases, the power density of illumination is maintained at 372 µWcm$^{-2}$. The current in all these photodetectors increases under illumination at every biasing voltage. A sharp increase of photo-induced current at zero bias can be attributed to the photo-voltaic effect. Device grown using target with 0.4 at.% of Si-content (D26) exhibits the maximum change in photo-induced current of 275 nA as can be seen in Figure 4(c). This demonstrates the self-powered operation of these devices.

Responsivity $R$ of a photodetector can be defined as $R = (I_p - I_d)/PA$, where $I_p$ the photo-induced current under illumination, $I_d$ the dark current, $P$ the power density of the incident light and $A$ the illuminated active area of the device.[40] Spectral responsivity profiles $R(\hbar\omega)$ recorded at zero bias for the set of devices are plotted in Figure 5. Evidently in all cases, the onset of responsivity appears at around 370 nm matching the band-gap of GaN. $R(\hbar\omega)$ remains to be finite down to the minimum wavelength of light illumination (240 nm) used in this study. It is interesting to note that the peak responsivity for the undoped sample appears at ~ 343 nm that is close to the GaN band-edge. On the other hand, for the devices with Si-doped Ga$_2$O$_3$ layers, $R(\hbar\omega)$ profiles maximize at ~ 260 nm, which is close to the band-gap of Ga$_2$O$_3$. Moreover, the value of the peak-responsivity is found to be much more in the Si-doped samples than in the undoped one. The maximum value of *R* is found to be 56.8 mA/W in the device D26 with the Si-concentration of $8 \times 10^{18}$ cm$^{-3}$. Note that the depletion width in the undoped Ga$_2$O$_3$ layer is supposed to be much more than that of the p-type GaN side. Appearance of the $R(\hbar\omega)$ peak near the GaN band-edge in the undoped sample can be attributed to much stronger depletion field in the GaN side than the Ga$_2$O$_3$ side, which might be resulting in larger photo-voltaic effect in the GaN side. Similarly, the shifting of the $R(\hbar\omega)$ peak towards the Ga$_2$O$_3$ band-edge in case of the Si-incorporated samples can be attributed to the enhancement of the depletion field in Ga$_2$O$_3$ side as a result of Si-doping. Another important figure of merit of a photodetector is the detectivity $(D) = R\sqrt{A/2eI_d}$, which measures how sensitive the photodetector is to weak signals. The peak detectivity at the zero bias can be estimated as $2.3 \times 10^{11}$, $1 \times 10^{12}$, $3 \times 10^{12}$, $5.5 \times 10^{11}$ Jones for the devices D7, D16, D26 and D33, respectively. In terms of both the responsivity and detectivity, our best performing devices are quite comparable with the reported data as can be seen in Table 2, which tabulates all the important figure of merits namely the responsivity, detectivity and the response times of the devices investigated here and the self-powered Ga$_2$O$_3$/GaN photodetectors reported in the literature by other groups.

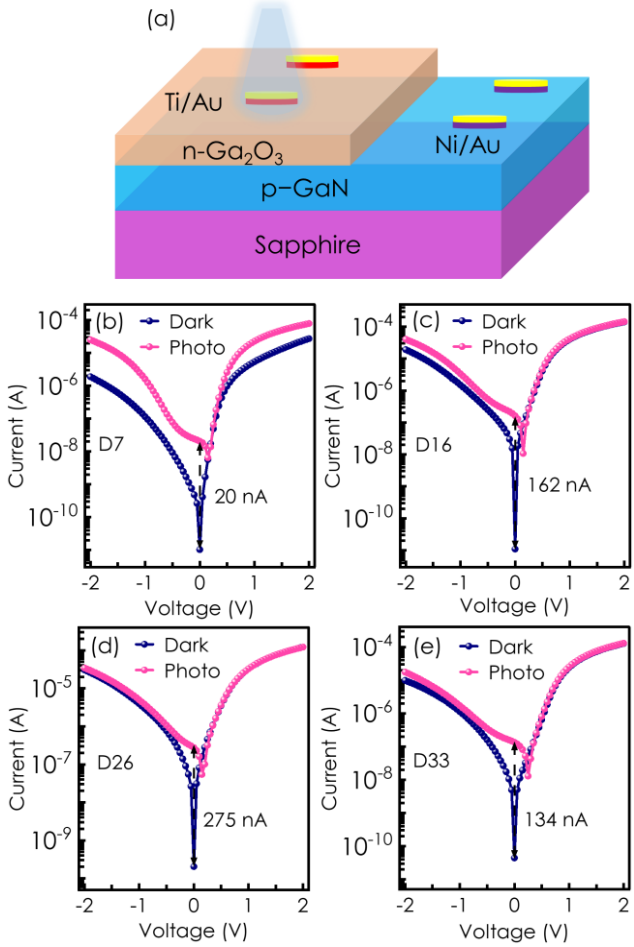

**FIG. 4.** (a) Schematic diagram of the device. *I–V* characteristics recorded for the device (b) D7, (c) D16, (d) D26 and (e) D33 under dark and 254 nm light illuminated with a power density of 372 μWcm$^{-2}$ conditions.

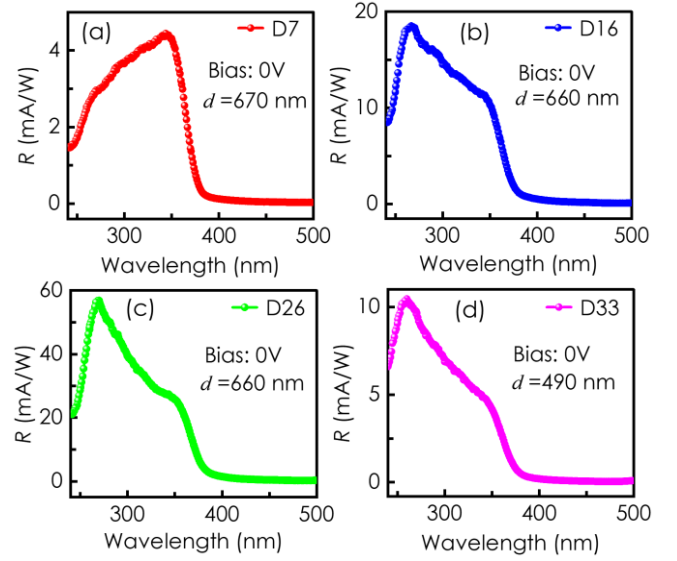

**FIG. 5.** Spectral responsivity profiles recorded at the zero bias condition for the devices (a) D7, (b) D16, (c) D26 and (d) D33.

Figure 6 plots the photocurrent recorded at the zero bias for the devices D7, D16, D26 and D33 under 254 nm light illumination as a function of the illumination power in the log-log scale. Except for the device D33, in all cases, the slope of the plots is estimated to be ~1. This shows that the dependence of the responsivity of these devices on the incident power is linear. In case of D33, where the Ga$_2$O$_3$ layer has the highest Si-content, the slope is obtained as ~1.4. The deviation of the slope from 1 in this case, might be due to the inclusion of higher density of defects and imperfections in the Ga$_2$O$_3$ layer at such a high density of Si-incorporation. One can also estimate the lowest detectable power (LDP) of a device by linearly extrapolating these plots to the dark noise floor level, which in the present case is ~20 pA. In all the devices, the LDP is estimated to be only a few nW.

In order to obtain further insight into the role of doping on the performance of these devices, in Figure 7, the integrated responsivity (ℝ) over the entire spectral range of study at zero bias is plotted as a function of the Ga$_2$O$_3$ layer thickness for every batch of Si-doped samples under the same illumination conditions. Please see Fig. S4 (supplementary material) for the spectral responsivity of the devices with Ga$_2$O$_3$ layer of different thicknesses and Si-concentrations. Evidently, in all cases, ℝ as a function of the thickness passes through a maximum. As mentioned before that apart from the layer-thickness, two other length-scales namely the penetration depth of the light (inverse of the absorption coefficient) and the depletion width become important in determining ℝ. Ideally, the responsivity for a light of a given wavelength (above band-gap) should be maximum when the three length-scales are equal. This is because the light energy can be most efficiently utilized at this condition. Since the absorption length-scale, which is typically ~100 nm in a semiconductor for the above band-gap light, is intrinsic to the materials, the maximum of ℝ is expected to be achieved when the thickness of the Ga$_2$O$_3$ layer matches with the depletion width. It is also evident from the figure that except for the device grown with the highest Si-content (device D33), the value of ℝ shows an increasing trend with the Si content. The rise of $\mathbb{R}$ with the Si-inclusion in the $Ga_2O_3$ layer can be explained in terms of the increase of the depletion field as a result of the enhancement of n-type doping level in the layer. However, all Si-ions might not be incorporated at the Ga substitutional sites to act as shallow donors in the sample. At a sufficiently high Si-concentration, it is plausible that a good fraction of the incorporated Si-ions are substituting the oxygen ions or going to the interstitial sites in the $Ga_2O_3$ lattice. These defects can produce acceptor like states and to some extent compensate the Si-donors. This might be the reason for the reduction of $\mathbb{R}$ in the sample with the highest Si-content. Another possible reason for the suppression of responsivity in this sample could be the reduction of carrier mobility due to the overall enhancement of the concentration of impurities, which act as scatters.

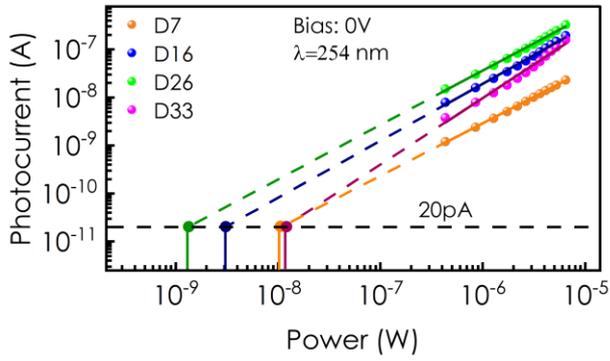

**FIG. 6.** Photocurrent (at zero bias) versus illumination power for 254 nm light in case of the devices D7, D16, D26 and D33. The lowest detectable powers for these devices are estimated by linearly extrapolating the plots (dashed lines) to the dark noise floor level.

The temporal response of the devices are studied in two different ways. In the first case, the photocurrent of the samples is recorded at zero bias as a function of time as the 254 nm illumination is switched on and off with a mechanical chopper. Temporal response of the photocurrent for the sample D26 are shown in Figure 8(a). The response profiles for the other three devices studied here are shown in the Figure S5(a) to (c) (supplementary material). In all cases, the rise and the decay profiles can be fitted with two exponential functions with different time constants, which range between a few ms to a few tens of ms. One can also define the rise($\tau_r$)[decay($\tau_d$)] time as the time required for the signal to reach $(1 - e^{-1})[e^{-1}]$ times the saturation value. Figure 8(b) compares the decay profiles for all the devices studied here. Evidently, the decay slows down with the increase of Si-content in the $Ga_2O_3$ layer. The growth-rate of the signal also shows similar trend. This is clearly visible in the Figure 8(c), where $\tau_r$ and $\tau_d$ are plotted as functions of the actual Si-content in these devices. This slowdown of the device response can be attributed to the generation of certain traps in the $Ga_2O_3$ layer due to Si-incorporation. These traps can offer a capture barrier to the photogenerated carriers trapped there, which makes their release taking longer time. This delay slows down the device response.

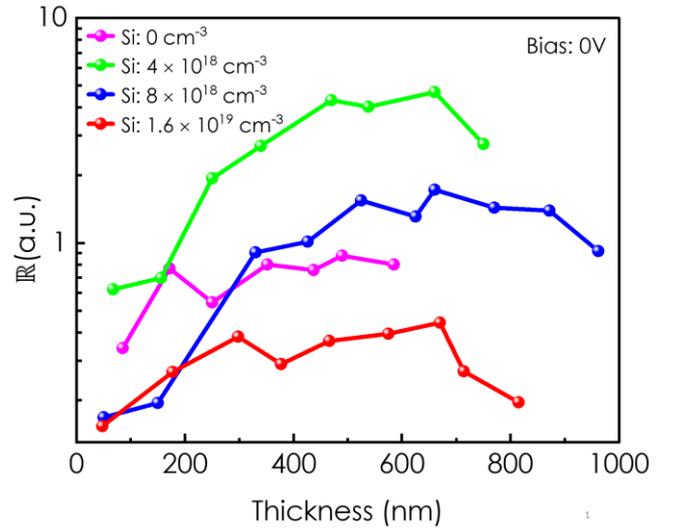

**FIG. 7.** Integrated responsivity $\mathbb{R}$ as a function of the thickness of the $Ga_2O_3$ film for the undoped and different Si-doped series of samples.

Figure 8(d) compares the normalized time response profiles for the device D33 recorded at different applied (reverse) biases, when the chopper switches on and off the 254 nm illumination. Evidently both the growth and decay rates slow down as the bias increases. Note that a similar bias dependence of the photoresponse rate has earlier been reported in n-ZnO/p-GaN heterojunction photodetectors, where it has been explained in terms of a competition between photo-voltaic (PV) and the persistent photoconductivity (PPC) effects.[40] At zero-bias condition, the photoresponse is governed only by the photo-voltaic (PV) effect, which is turned-on/vanished as soon as the light is switched on/off. This results in a fast photo response of the device. On the other hand, a reverse bias can setup a leakage current of the photogenerated carriers in the circuit, which may lead to PPC effect as a result of capture of the carriers by certain defect sites. As the bias increases, the PPC component in the overall photoresponse increases as compared to the PV component, which can slow down the response.[41] We believe that a similar picture can explain the observations of Figure 8(d) in these devices as well. Note that the response time of the devices can be altered by several orders of magnitude by applying bias. As far as neuromorphic device applications are concerned, such a large scale electrical tunability of photoresponse time is highly interesting. In order to check the stability and repeatability these devices are subjected to repeated on/off illumination cycles (0.7/0.7 s on/off time) over more than an hour. All devices show stable and repeatable performance over such a long period (see Fig. S6 in the supplementary material).

In order to explore whether the photo-response of these devices has any ultrafast component or not, a 266 nm picosecond pulse laser with pulse width of 250 ps is used for illumination. Normalized photocurrent $I_{ph}^N$ versus time profiles are plotted for the reference sample D7 and one of the Si-containing samples D26 in Figure 9. Inset of the figure shows $\ln(I_{ph}^N)$ versus time plots for the data. Time-response profiles for other devices (D16 and D33) are presented in the Figure S4d (Supporting Information). As evident from the figures, the rise-time is much faster than the decay-time. In fact, in all cases, the rise-time is found to be about 1 ns. As shown in the inset of Figure 9, the decay part can be fitted with single straight line in case of the reference undoped sample D7. While, the profile for the Si-containing sample D26 clearly has two slopes. Interestingly, the faster component has the same slope as that of the reference sample. The fast and the slow decay components are found to be ~26 and ~80 ns, respectively. The origin of the slower decay component might be certain trap centres generated due to Si-incorporation. The ~80 ns decay component has also been observed in other Si- doped samples as well (see Fig. S5(d) in the supplementary material). It is noticeable in Table 2 that our best performing device has the fastest response time reported so far in this type of devices.

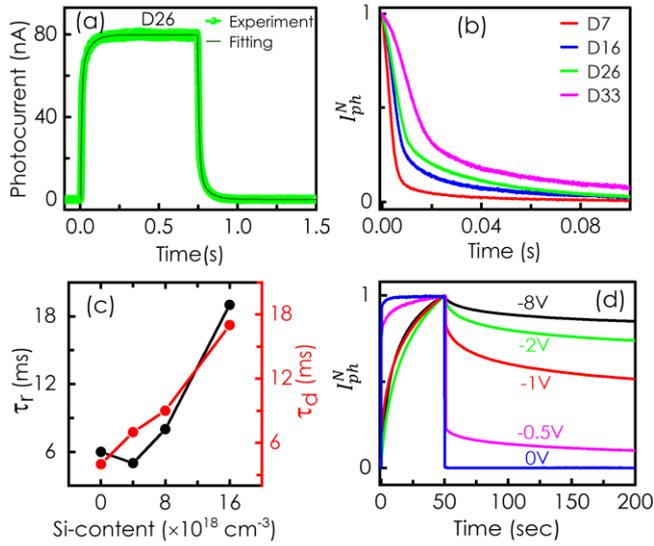

**FIG. 8.** (a)Time response profiles at zero bias for the device D26 as the illumination with 254 nm light of power density of ∼67 μWcm$^{-2}$ is switched on and off using a mechanical chopper. (b) The normalized decay profiles for the different devices from the experiments carried out in the same way as in (a). (c) Plots of the rise ($\tau_r$) and decay ($\tau_d$) times as functions of the Si-concentration. (d) Time response profiles of D33 under different reverse bias after switching on and off 254 nm light. Normalization is done after subtracting the dark current from the maximum photocurrent.

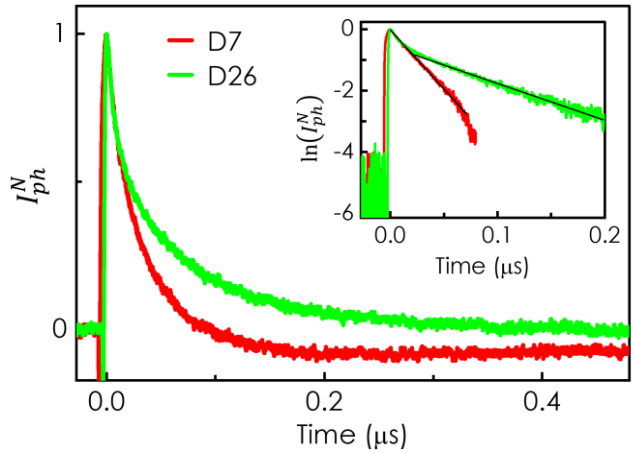

**FIG. 9.** $I_{ph}^N$ versus time profiles at zero bias for the device D7 and D26 when illuminated with a 266 nm picosecond pulsed laser with pulse width of 250 ps. Inset: $ln(I_{ph}^N)$ versus time plots for the two devices.

### IV.    Conclusion

Si-doped ($\bar{2}$01) β-Ga$_2$O$_3$ epitaxial layers are grown by PLD technique on p-type c-GaN/sapphire substrates. The current-voltage characteristics of these heterojunction devices show rectifying behavior establishing their p-n junction nature. These devices are found to perform as highly efficient self-powered visible blind UV photodetectors. It has been observed that the best performance can be achieved at certain optimum Ga$_2$O$_3$ layer thickness and Si-concentration. The maximum responsivity and detectivity are obtained as 56.8 mA/W and $3 \times 10^{12}$ Jones, respectively. The lowest power detectable by these devices is found to be  as low as a few nW.  Further, photoresponse time is found to be only a few tens of nanoseconds that enhances the potential of these devices for applications in ultrafast UV detection. These detectors also exhibit a slower time-scale for photoresponse, which is a few tens of milliseconds. Moreover, the response time is found to be increased by several orders of magnitude through application of reverse bias. Such an electrical tuneability of the response time of these devices could be potentially interesting for their neuromorphic device applications.

Table 2. Comparison of performance of $Ga_2O_3$/p-GaN based self-powered photodetectors reported in the literatures

| PD structure | Illumination wavelength $\lambda$[nm] | Responsivity $R$ [mA/W] | Rise/decay times [ms] | Detectivity $D$ [Jones] | Refs. |
|---|---|---|---|---|---|
| $Ga_2O_3$/p-GaN | 254 | 28.44 | 140/70 | $1.23 \times 10^{11}$ | 23 |
| Sn:$Ga_2O_3$/p-GaN | 254 | $3.05 \times 10^3$ | /18 | - | 19 |
| $Ga_2O_3$/GaN | 266 | 1.2 | 980/2000 | $1.35 \times 10^{10}$ | 42 |
| Nanoporous $Ga_2O_3$/GaN | 254 | 43.9 | 630/480 | $2.7 \times 10^{11}$ | 43 |
| Ta:$Ga_2O_3$/p-GaN(porous) | 222 | 35.4 | 410/340 | - | 30 |
| Ta:$Ga_2O_3$/ i-$Ga_2O_3$/p-GaN | 222 | $8.67 \times 10^3$ | 86/50 | $1.08 \times 10^{14}$ | 38 |
| Ta:$Ga_2O_3$/p-GaN | 254 | 50 | 350/230 | $6 \times 10^{11}$ | 44 |
| Si:$Ga_2O_3$ /i-$Ga_2O_3$/p-GaN | 260 | 72 | 7/19 | $3.22 \times 10^{12}$ | 17 |
| Si:$Ga_2O_3$/p-GaN | 260 | 14 | 37/42 | $5.31 \times 10^{11}$ | 17 |
| n-$Ga_2O_3$/p-GaN | 264 | 250 | /0.031 (@260 nm) | $3.3 \times 10^{12}$ | 7 |
| a-$Ga_2O_3$/p-GaN | 254 | $5.65 \times 10^3$ | 260/410 | $8.46 \times 10^{12}$ | 37 |
| n-$Ga_2O_3$/p-GaN | 254 | 44.98 | 383/96 | $5.33 \times 10^{11}$ | 45 |
| Si:$Ga_2O_3$/p-GaN | 270<br>254<br>266 | 56.8<br>36.5 | <br>6/10<br>1 ns/80 ns | $3 \times 10^{12}$<br>$1.9 \times 10^{12}$ | This work |

**Acknowledgment**

We acknowledge the financial support provided by the Science and Engineering Research Board (SERB), under Grant No. CRG/2022/001852, Government of India. Mr. Ajoy Biswas would like to thank University Grants Commission (U.G.C), Government of India for the fellowship. We would like to thank Sophisticated Analytical Instrument Facility (SAIF), Industrial Research and Consultancy Centre (IRCC), and the Centre for Excellence in Nanoelectronics (CEN), IIT Bombay for the use of various facilities. We are sincerely thankful to UGC-DAE CSR, Indore for SIMS facility. Additionally, authors would like to thank Mr. Sounak Samanta and Ms. Sana Ayyuby for their help in the wire bonding of the devices.

**Conflict of Interest**

The authors declare no conflict of interest

**Data Availability Statement**

Data that support the findings of this study are available from the corresponding author upon reasonable request.

**Keywords**

Epitaxial β-$Ga_2O_3$ film, Doping, Thickness, Self-powered Photodetector, Responsivity, Fast response, Tuneable response time.

# Supplementary Material for

Si-Ga$_2$O$_3$/p-GaN epitaxial heterostructure based self-powered and visible-blind UV photodetectors with fast and electrically tuneable response time

Ajoy Biswas, Amandeep Kaur, Bhabani Prasad Sahu, Sushantika Saha, Umakanta Patra, Pradeep Sarin, Subhabrata Dhar*

Department of Physics, Indian Institute of Technology Bombay, Powai, Mumbai 400076, India

*Email: dhar@phy.iitb.ac.in

## S1. Thickness and Si-concentration of the Ga$_2$O$_3$ layers

**Table S1**

| Batch 1 (undoped Ga$_2$O$_3$) | | Batch 2 (Si: $4 \times 10^{18}$ cm$^{-3}$ in Ga$_2$O$_3$) | | Batch 3 (Si: $8 \times 10^{18}$ cm$^{-3}$ in Ga$_2$O$_3$) | | Batch 4 (Si: $1.6 \times 10^{19}$ cm$^{-3}$ in Ga$_2$O$_3$) | |
|---|---|---|---|---|---|---|---|
| Devices | Thickness of Ga$_2$O$_3$ film (nm) | Devices | Thickness of Ga$_2$O$_3$ film (nm) | Devices | Thickness of Ga$_2$O$_3$ film (nm) | Devices | Thickness of Ga$_2$O$_3$ film (nm) |
| D1 | 48 | D10 | 50 | D20 | 68 | D28 | 85 |
| D2 | 177 | D11 | 150 | D21 | 156 | D29 | 172 |
| D3 | 298 | D12 | 330 | D22 | 250 | D30 | 250 |
| D4 | 377 | D13 | 426 | D23 | 340 | D31 | 352 |
| D5 | 466 | D14 | 525 | D24 | 470 | D32 | 437 |
| D6 | 575 | D15 | 625 | D25 | 538 | D33 | 490 |
| D7 | 670 | D16 | 660 | D26 | 660 | D34 | 585 |
| D8 | 714 | D17 | 770 | D27 | 750 | | |
| D9 | 815 | D18 | 872 | | | | |
| | | D19 | 962 | | | | |

## S2. *I-V* characteristics

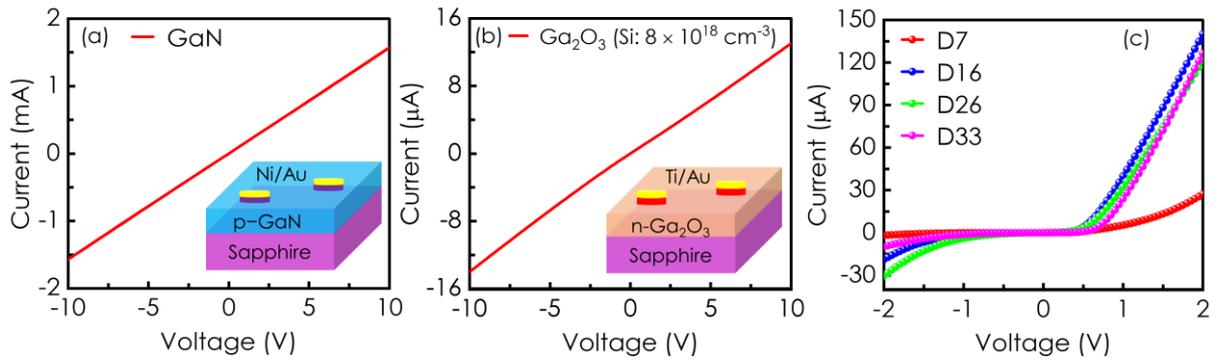

**Fig. S2.** Current-voltage (*I-V*) characteristics of the contacts on (a) GaN template and (b) a Si-doped $Ga_2O_3$ film grown on c-sapphire substrate. (c) I-V characteristics measured between the contacts in GaN and $Ga_2O_3$ sides for the devices studied here.

## S3. SEM Surface images

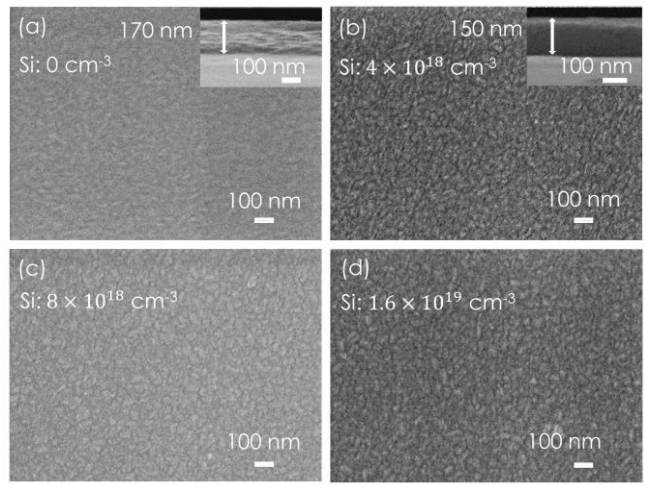

**Fig. S3.** SEM surface images of $Ga_2O_3$ layer of Si-concentration (a) 0 [device D2], (b) $4 \times 10^{18}$ cm$^{-3}$ [device D11], (c) $8 \times 10^{18}$ cm$^{-3}$ [device D21] and (d) $1.6 \times 10^{19}$ cm$^{-3}$ [device D29]. Insets of (a) and (b) show the cross-sectional images. Thickness of the $Ga_2O_3$ layer in these devices can be found in Table S1.

## S4. Responsivity of four different series photodetector devices

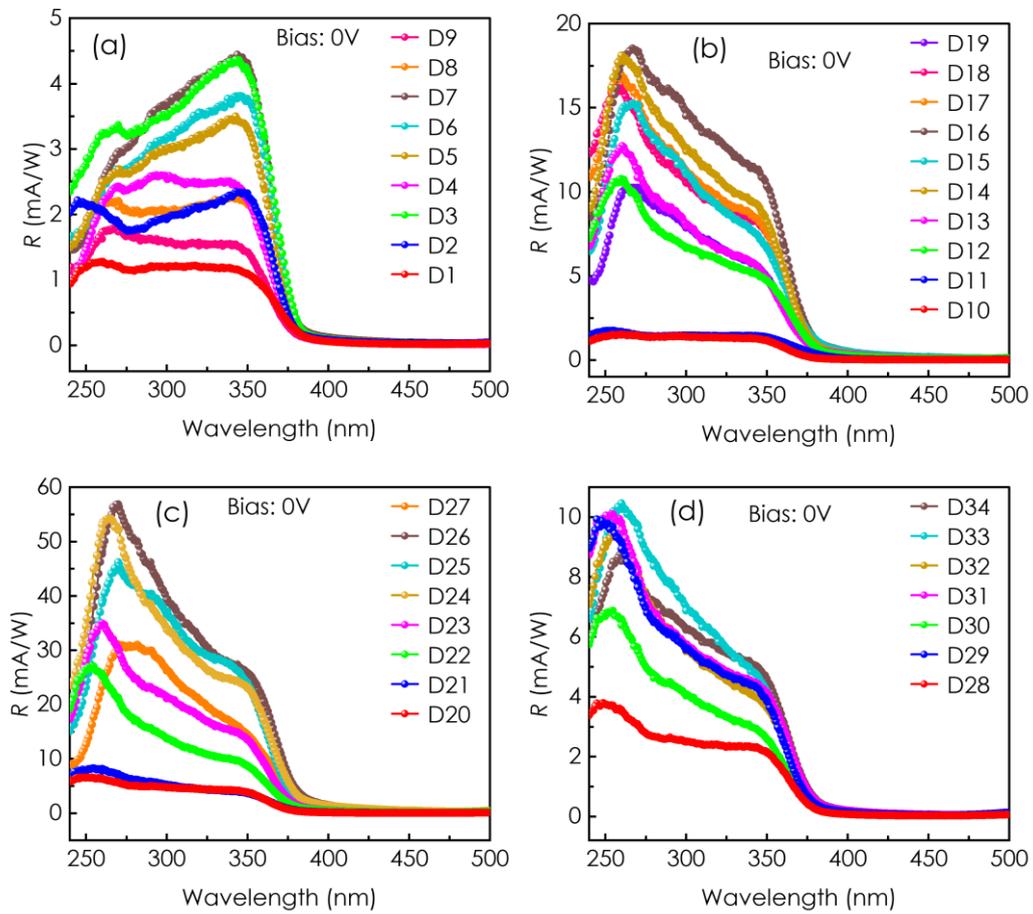

**Fig. S4**. Spectral responsivity for devices of (a) Batch 1, (b) Batch 2, (c) Batch 3 and (d) Batch 4. Information about the $Ga_2O_3$ layer in different devices can be found in Table S1.

## S5. Time response

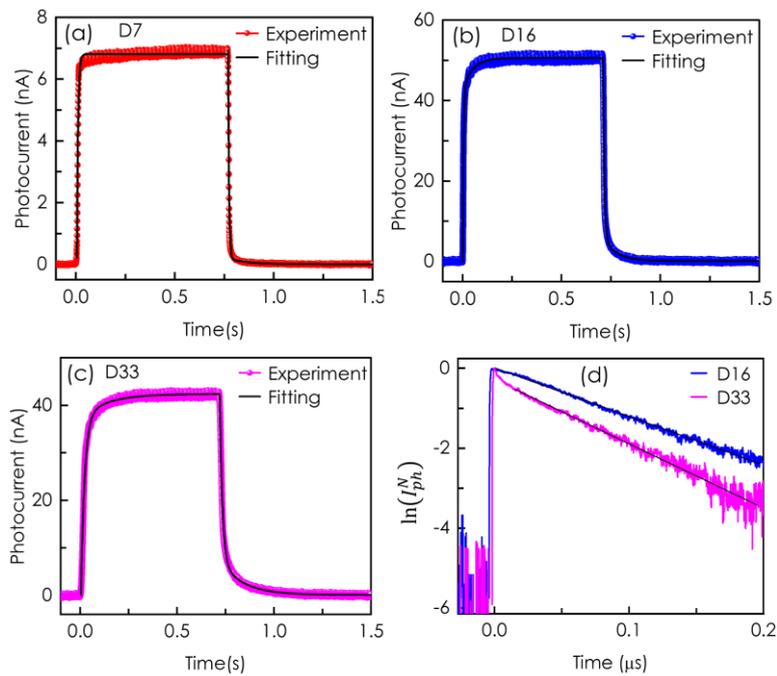

**Fig. S5.** Time response profiles recorded at zero bias condition for the device (a) D7, (b) D16 and (c) D33 as the illumination with 254 nm light of power density of ~67 μWcm$^{-2}$ is switched on and off using a mechanical chopper. (d) $\ln(I_{ph}^{N})$ versus time plots for device D16 and D33 when illuminated with a 266 nm picosecond pulsed laser with pulse width of 250 ps.

## S6. Stability

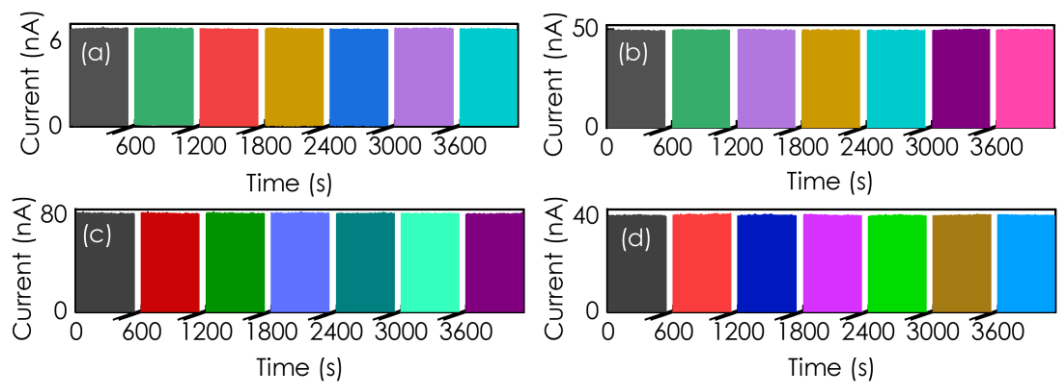

**Fig. S6.** Time variation of the photocurrent measured at zero bias when the device is subjected to repeated on/off cycles of illumination (0.7/0.7 s on/off time) with 254 nm light over a long period for the device (a) D7, (b) D16, (c) D26 and (d) D33.